\documentstyle[twocolumn,prl,aps]{revtex}
\begin{document}
\draft
\title{Collective excitation frequencies and  damping rates of 
a two-dimensional deformed trapped Bose gas above the critical 
temperature} 
\author{Tarun Kanti Ghosh }
\address
{The Institute of Mathematical Sciences, C. I. T. Campus. Chennai- 
600 113, India.} 
\date{\today}
\maketitle

\begin{abstract}
We  derive a  equation of motion for the velocity fluctuations of a  2D deformed 
trapped Bose gas  just above the critical temperature in the hydrodynamical 
regime. 
From this equation, we calculate the eigenfrequencies and the corresponding 
density fluctuations for a few low-lying excitation modes. 
Using the method of averages, we derive a  
dispersion relation in a deformed trap at very high temperature that interpolates 
between the 
collisionless and hydrodynamic regimes. We make use of this dispersion relation
to calculate the frequencies and the damping rates for monopole and quadrupole mode 
in both the regimes. We also discuss the time evolution of the wave packet width  of
a Bose gas in a time independent as well as time dependent trap.

\end{abstract}


\narrowtext
 
\section{introduction}
There has been renewed interest in the Bose-Einstein condensation(BEC)
after its experimental demonstration by Anderson et.al \cite{anderson} in
a magnetically trapped rubidium  gas. In other experiments the
Bose-Einstein condensate of trapped lithium \cite{bradley} and sodium
vapor \cite{davis} is also observed. There has been much interest 
in the theoretical understanding of this system \cite{dal}. 
 The low-lying collective excitations above the critical temperature in 
the hydrodynamic region has been discussed by Griffin et.al \cite{griffin} using 
the kinetic theory. Damping of the hydrodynamic modes in a trapped Bose gas above the 
Bose-Einstein critical temperature ($ T_c $) is also discussed by Odelin et.al 
\cite{odelin} and Kavoulakis et.al \cite{pethick}. Frequencies of the low-lying 
excitation modes at $ T = 0 $ have been discussed by S. Stringari \cite{strin}.

 After the discovery of BEC in a trapped alkali atom, the influences of the
dimension of a Bose systems has been a subject of extensive studies 
 \cite{petrov}. In our present technology one can
frezze the motion of the trapped particles in one direction to create a quasi-2D Bose 
gas. In the frozen direction the particles executes the zero point motion. To 
achive this quasi-2D system, the frequency in the frozen direction should be
much larger than the frequency in the X-Y plane and the mean field
interactions among the particle. It has been shown by V. Baganato et.al 
\cite{baganato} that for an ideal 2D Bose gas under harmonic trap a macroscopic
 ocupation of the ground state can exist at temperature
$ T < T_c = \sqrt{\frac{N \lambda}{\zeta(2)}}\frac{\hbar \omega}{k_{B}} $.
$ \lambda = \frac{\omega_y}{\omega_x}$ is the deformation parameter.
Some experiment shows the possibility of creating a quasi-2D trapped 
Bose gas \cite{gauck}. 

It is well known that the excitation frequencies for monopole and quadrupole modes are 
$ 2 \omega_0 $ and $ \sqrt 2 \omega_0 $ respectively in a 2D isotropic trapped Bose gas. Using the 
approximation, $ \omega_z >> \omega $, the dispersion relation of the 
excitation frequencies \cite{griffin}, \cite{odelin} does not produce the correct 
frequencies for monopole and quadrupole modes in a 2D trapped Bose system.
There has been no systematic study on the collective excitations
of a 2D deformed trapped Bose gas above the critical temperature.

The purpose of this paper is to give analytic results for the dispersion law 
of low-lying collective modes in 2D deformed trapped Bose gas  
and their damping rates in both regime, hydrodynamic and collisionless.

Above the critical temperature, one can distinguish two regimes, the
hydrodynamic(collisional) one where collisions ensure the local thermal
equlibrium and collisionless where the motion is described by the single
particle hamiltonian. In the hydrodynamic region, the characteristic mode
frequency is small compared to the collision frequecy ($ \omega \tau << 1 )
$. In the collisionless region ($ \omega \tau >> 1 $ ), the collision are
not important.

The paper is organised as follows. We derive in sec[II] a closed equation of motion for 
the velocity fluctuations of a 2D deformed trapped Bose gas just above the critical
 temperature ($ T>T_{c}$) using the kinetic theory. We make use of this 
equation in 
sec[III] to calculate the excitation frequencies for a few low-lying collective 
modes and the corresponding density fluctuations. In sec[IV], we derive a 
dispersion relation in a 2D deformed trap at very high temperature using the method of 
averages that interpolates between the collisionless and hydrodynamic regimes. From 
this dispersion relation, we calculate the eigenfrequencies and damping rates for
monopole and quadrupole mode. We discuss the evolution of
the wave packet width of a Bose gas in a time independent as well as time dependent
trap. In sec[V], we presents a summary of our work.

\section{Hydrodynamic Equation of motion for the velocity fluctuations}
We shall discuss the collective modes of a 2D deformed trapped Bose
gas in the hydrodynamic regime just above the critical temperature $ T > T_{c} $ 
using the kinetic theory. In the low energy excitations, we can use the
semiclassical approximation for the dynamics of a Bose gas, using
the following Boltzmann equation \cite{haung} for the phase space distribution 
function
 $ f(\vec{r},\vec{p},t) $
\begin{equation} \label{boltz}
\frac{\partial f}{\partial t} + \vec
v.\nabla_{\vec r}f + \frac{\vec F}{ m} . \nabla_{\vec v}f = I_{coll}(f)
\end{equation}
 where $ I_{coll} $ is the collisional integral and $ \vec F = -\nabla U_{0}(\vec
r) $. The trap potential is $ U_0 (r) = \frac{1}{2} m(\omega_{x}^2 x^2 +
\omega_{y}^2 y^2 ) $.

 In the hydrodynamic regime, collisions ensures the local thermodynamic
equilibrium.
  To lowest order, the perturbed distribution function produced by a slowly varying
external potential is the equlibrium Bose distribution function
\begin{equation} 
f(\vec{r},\vec{p},t) = [\exp(\beta(\vec{r},t)\eta(\vec{r},t) - 1]^{-1}
\end{equation}
\begin{equation}
\eta(\vec{r},t) =\frac{[\vec{p} - m\vec{v(\vec{r},t)}]^{2}}{2m} -
\mu (\vec{r},t)
\end{equation}
$ \mu (\vec{r},t) $ is the chemical potential. The inverse temperature is $ \beta 
(\vec{r},t)  = \frac{1}{k_{B}T(\vec{r},t)} $.
We are only interested in a small perturbations around the equilibrium states.

The conservation laws are \cite{baym}:
\begin{equation}\label{density}
\frac{\partial n(\vec{r},t)}{\partial{t}} +
\nabla . [n_{0}(\vec{r})\delta\vec{v}(\vec{r},t)] = 0    
\end{equation}
\begin{equation}\label{velocity}
m n_{0}(\vec{r})\frac{\partial \delta{\vec{v}}}{\partial t} =
-[\nabla P(\vec{r},t)+n(\vec{r},t) \nabla U_{0}(\vec{r})] 
\end{equation}
\begin{equation}\label{energy}
\frac{\partial E(\vec{r},t)}{\partial t}=-\nabla.[(P(\vec r ) + E(\vec r
))\delta
\vec{v}] - n_{0} \delta \vec{v}.\nabla U_{0}(\vec{r})
\end{equation}
These conservation laws are obtained from Eq. (\ref{boltz}) multiplying by 1, $ \vec p $,
$ \frac{\vec p^2}{2m} $ and integrating the resulting equation over $ \vec p $. 
During  collisions, the total number of particles N, momentum $ \vec p $, and energy
$ \frac{\vec p^2}{2m} $ are conserved, so the collisional term vanishes.
$ \delta {\vec v} $ is the velocity fluctuation around the equilibrium states.

Using the quantum statistical mechanics, pressure and density can be written as 
\begin{equation}
\frac{P}{ k_{B} T} = \frac{g_{2}(z)}{ \Lambda^2 }
\end{equation}
\begin{equation}
n = \frac{g_{1}(z)}{ \Lambda^2 }
\end{equation}
where  $ g_{n}(z) = \sum_{i=1}^{\infty} (\frac{z^i}{i^n}) $ are the Bose-Einstein 
functions.
 $ z(r,t) = e^{\beta(r,t) \mu(r,t)} $ is the local thermodynamic fugacity which 
is always less than one.
$ \Lambda = \sqrt{\frac{ 2\pi \hbar^2}{m k_{b} T }} $ is the 
thermal de-Broglie wave length.

One can easily get the relation,
\begin{equation}\label{pressure}
 P(\vec{r},t) = E(\vec{r},t)
\end{equation} 
 in 2D .
Using Eq. (\ref{pressure}), Eq. (\ref{energy}) can be written as 
\begin{equation}\label{pre}
\frac{\partial P(\vec{r},t)}{\partial t}=- 2 \nabla.[(P_{0}(\vec r ) \delta \vec{v}(\vec r,t)]
 - n_{0} \delta \vec{v}(\vec r,t).\nabla U_{0}(\vec{r})
\end{equation}

Taking the time derivative of Eq. (\ref{velocity}) and using Eqs. (\ref{density}) and 
(\ref{pre}), we get
\begin{equation}\label{main}
m\frac{\partial^{2}\delta\vec{v}}{\partial t^{2}}=2\frac{P_{0}(\vec{r})}
{n_{0}(\vec{r})}\nabla [\nabla . \delta \vec{v}]-[\nabla 
. \delta \vec{v}] \nabla U_{0}(\vec{r}) - 
\nabla [\delta \vec{v} . \nabla U_{0}(\vec{r})]
\end{equation}
The closed equation of motion for the velocity fluctuations has  been
derived by Griffin et.al  \cite{griffin} for 3D trapped  Bose system.
The term $ \frac{P_{0}(\vec r)}{n_{0}(\vec r)} $ of 
(\ref{main}) is associated with the Bose statistics.

Without any external potential $ U_{0}= 0 $, the Eq. (\ref{main}) becomes
\begin{equation}
m\frac{\partial^{2}\delta\vec{v}}{\partial t^{2}}=2\frac{P_{0}(\vec{r})}
{n_{0}(\vec{r})} \nabla[\nabla . \delta \vec{v}]
\end{equation}
It has the plane wave
solution with the dispersion relation $\omega^{2}=c^{2} k^{2} $. The sound
velocity is $c^{2}= \frac{2 P_{0}(\vec{r})}{ m n_{0}(\vec{r})}$
or $c^{2}=\frac{2 k_{B} T_{0} g_{2}(z_0)}{m g_{1}(z_0)} $
where $ z_0 = e^{\frac{ \mu_0 (r)}{ k_{B} T_0 }} $ and $ \mu_0 (r) = \mu - U_0 (r) $.
At high temperature ($  z <<1 $ ), the sound velocity becomes $ c^{2}=\frac{2 k_{B}
T_{0} }{m } $. This sound velocity exactly matches with  known 
result.
 
From the continuity Eq. (\ref{density}) we have,
\begin{equation}
\frac{\partial \delta n(\vec r,t) }{\partial t } = - (\nabla . \delta
\vec v ) n_{0}(\vec r) - \delta \vec v(\vec r,t) .\nabla n_{0}(\vec r,t)
\end{equation}
The density fluctuation is given by $ \delta n(\vec r,t) = \delta n(\vec r
) e^{- i \omega t} $.
In classical limit, the static density profile is
$ n_{0}(\vec r) = n_{0}(\vec r = 0 ) e^{- \frac{ m (\omega_{x}^2 x^2 + 
\omega_{y}^2 y^2) }{2 \theta }} $ , where $ \theta = k_{B} T $.

\section{ Eigenfrequencies and The corresponding density fluctuations in the 
hydrodynamic regime }
1) The normal mode solution of (\ref{main}) is $ \delta \vec v(\vec r) = 
\nabla (z^{l}) $, here
$ z =( x+iy) $ and $ l > 0 $. The excitation frequencies and the associated 
density fluctuations are $\omega^2 = l \omega_{x}^2 $ ,
$\delta n_{x} \sim \omega_{x}^2 x z^{(l-1)} n_0(\vec r)$
and $\omega^2 = \omega_{x}^2 + ( l- 1) \omega_{y}^2 $  ,
$ \delta n_{y} \sim \omega_{y}^2 y z^{(l-1)} n_0 (\vec r)$.

For isotropic trap, the frequency is
 $ \omega= \omega_{0} \sqrt l $.
The corresponding density fluctuation is 
$ \delta n(\vec r) \sim  n_{0}(\vec r ) r^l e^{ i l \theta } $.
At $ r = 0 $ there is no density fluctuation. There is a maximum
density fluctuation at $ r = \sqrt{\frac{l \theta }{m \omega_{0}^2}} $.

2) The other solution of Eq. (\ref{main}) is $ \delta v(\vec r) =
\nabla[\alpha x^{2} \pm \beta y^{2}] $. The positive sign is for the monopole mode 
and 
the negative sign is for quadrupole mode. In a deformed trap, the excitation 
frequencies are
\begin{equation}\label{mono}
 \omega^{2} = \frac{1}{2}[3(\omega_{x}^{2} + \omega_{y}^{2})\pm
\sqrt {9(\omega_{x}^{2} + \omega_{y}^{2})^{2} - 32 \omega_{x}^{2}
\omega_{y}^{2}}]
\end{equation}
For an isotropic tarp, it becomes
$ \omega = 2\omega_{0} $ or $ \omega = \sqrt 2 \omega_{0} $. Hence
in the anisotropic trap the monopole mode is coupled to the quadrupole mode.
If $ \omega_{x} << \omega_{y} $, the lowest excitation frequency 
 is $ \omega = \sqrt {\frac{8}{3}} \omega_{x} $.
If $ \omega_{x} >> \omega_{y} $, the lowest excitation frequency
is $ \omega = \sqrt {\frac{8}{3}} \omega_{y} $.
The density fluctuation for the monopole mode is 
$ \delta n(\vec r) \sim  [ 2 - \frac{m (\omega_{x}^2 x^2 + \omega_{y}^2 y^2)}
{\theta } ] n_{0}(\vec r) $ where as the density fluctuation for quadrupole 
mode is $ \delta n( \vec r) \sim  (\omega_{y}^2 y^2 - \omega_{x}^2 x^2 )
 n_{0}( \vec r ) $.

3) There is another quadrupole mode which has velocity field 
$ \delta \vec v (\vec r) = \nabla (xy) $. This is also called scissors mode 
\cite{ode}.
The excitation frequency is  $ \omega^{2} = \omega_{x}^{2} + \omega_{y}^{2} $
and the corresponding density fluctuation is $ \delta n(\vec r ) \sim 
(\omega_{x}^2 + \omega_{y}^2 ) x y n_{0}(\vec r) $.
In isotropic trap $ \omega^2 = 2 \omega_{0}^2 $, which agrees with that for  the
scissors mode in hydrodynamic regime above $ T_{c} $ \cite{ode}.

\section{Method of Averages}
At very high temperature, the dynamical behaviour of a dilute gas is described by the 
Boltzmann tarnsport equation. Here we include the collisional term in the Boltzmann
transport equation and study the eigenfrequencies for 
monopole  and quadrupole mode using the method of averages \cite{odelin}.
These two modes are coupled in a deformed trap.

From Eq. (\ref{boltz}), one can get the equations for the average of a 
dynamical quantity  $ \chi(\vec r,\vec v) $ is {\cite{haung}, \cite{ford}}
\begin{equation}\label{boltzman}
\frac{d<\chi>}{dt} - <\vec v .\nabla_{\vec r}\chi > - < \frac{\vec F}{
m} . \nabla_{\vec v} \chi > = <I_{coll} \chi >
\end{equation}
where the average is taken in phase space and $ <\chi> $ can be written as
\begin{equation}
< \chi > = \frac{1}{N} \int d^2r d^2v f(\vec r,\vec v,t) \chi(\vec r,\vec
v)
\end{equation}
$ <\chi I_{coll} > $ can be defined as
\begin{equation}
<\chi I_{coll} > = \frac{1}{4 N}\int d^2r d^2v [\chi_{1} + \chi_{2} -
\chi_{1}^{\prime } - \chi_{2}^{\prime } ]  I_{coll}(f)
\end{equation}
If $ \chi = a(\vec r ) + \vec b(\vec r).\vec v +c(\vec r) \vec v^2 $
, for elastic collision the collisional term is zero 
\cite{odelin}, \cite{haung}.
a, $ \vec b $, and c are arbitrary functions of the position. 

Now we define the following quantities,
\begin{equation}
\chi_{1} = x^2 +  y^2
\end{equation}
\begin{equation}
\chi_{2} = y^2 - x^2
\end{equation}
\begin{equation}
\chi_{3} = x v_{x} + y v_{y}
\end{equation}
\begin{equation}
\chi_{4} = y v_{y} - x v_{x}
\end{equation} 
\begin{equation}
\chi_{5} = v_{x}^2 + v_{y}^2
\end{equation}
\begin{equation}
\chi_{6} = v_{y}^2 -  v_{x}^2
\end{equation}
Using the Boltzmann kinetic equation (\ref{boltzman}), we get the following
closed  set of equations.
\begin{equation}\label{bequ}
<\ddot \chi_{1} > = 2 <\chi_{5} > - t <\chi_{1} > + \epsilon < \chi_{2} >
\end{equation}
\begin{equation}
<\ddot \chi_{2} > = 2 <\chi_{6} > - t <\chi_{2} > + \epsilon < \chi_{1} >
\end{equation}
\begin{equation}
<\ddot \chi_{3} > = 2 \epsilon <\chi_{4} > - 2 t <\chi_{3} > 
\end{equation}  
\begin{equation}
<\ddot \chi_{4} > = 2 \epsilon <\chi_{3} > - 2 t <\chi_{4} > 
-\frac{<\chi_{6}>}{\tau}
\end{equation}
\begin{eqnarray}
<\ddot \chi_{5} > & = &  \epsilon <\chi_{6} > -  t <\chi_{5} > - \epsilon t <
\chi_{2}> \\ \nonumber & + & \frac{\epsilon^2 + t^2}{2} < \chi_{1} >
\end{eqnarray}
\begin{eqnarray}\label{eequ}
<\ddot \chi_{6} > & = & -\epsilon t < \chi_{1} > +  \frac{\epsilon^2 + t^2}{2}
<
\chi_{2}> \\ \nonumber & + & \epsilon <\chi_{5} >  -  \frac{<\dot \chi_{6}>}{\tau}
 - t<\chi_{6} > 
\end{eqnarray}
where  double dot indicates the double derivative with respect to 
time.
$ t = \omega_{x}^2 + \omega_{y}^2 $ and
$ \epsilon = \omega_{x}^2 - \omega_{y}^2 $.
The $ \chi_{6} $ is not a conserved quantity, so the collisional
contribution comes only through the  $ \chi_{6} $ term.
We have used the fact that $ < \chi_{6} I_{coll} > = - \frac{\chi_{6}}{\tau
} $, where $ \tau $ is the relaxation time. This relaxation time $ \tau $
can be computed by a gaussian ansatz for the distribution function.
The relaxation time $ \tau $ is order of the inverse of the collision rate  
$\gamma_{coll} \sim  n(0) v_{th} \sigma_{0} $, where $ v_{th} = 
\sqrt{\frac{ \pi k_{B} T }{2 m }} $ is the mean thermal velocity and
$ n(0) = \frac{N m \omega_{0}^{2} \lambda}{2 \pi k_{B} T a_{z}} $ is the central 
density for a quasi-2D system. $ a_z $ is the osscilator length in the z-direction. 
Hence $ \tau \sim 
\frac{4 a_{z}}{N \sigma_0 \omega_{0}^{2} \lambda } \sqrt{ \frac{2 \pi k_{B} T}{ 
m}} $.  $ \sigma_{0} = 8\pi a^2 $ is the 3-D scattering cross-section. It  can be 
written in terms of  $ T_{c} $ as 
\begin{equation}\label{tau}
\tau \sim \sqrt{\frac{h}{m \pi^2 \sqrt{\zeta (2)}}} \frac{1}{(N 
\lambda)^{\frac{3}{4}}} (\frac{a_z}{a^2 \omega_{0}^{3/2}}) \sqrt{\frac{T}{T_c}}
\end{equation}
The relaxation time $ \tau $ varies as $ \sqrt T $ in a quasi-2D where as in 3D it 
varies as T \cite{pethick} .
Now we are looking for a solutions of Eqs. (\ref{bequ})-(\ref{eequ}) as $ e^{- i \omega t } $.
 We have the following dispersion relation
\begin{eqnarray}\label{inter}
(\omega^2 - 4 \omega_{x}^2 )(\omega^2 - 4 \omega_{y}^2 ) +
\frac{ i }{\omega \tau } [ \omega^4 - 3 \omega^2 (\omega_{x}^2 + \omega_{y}^2 ) +
\\ \nonumber 8 \omega_{x}^2 
\omega_{y}^2 ] & = & 0
\end{eqnarray}
This dispersion relation interpolates between the collisionless and hydrodynamic 
regimes.  In the hydrodynamical regime ( $ \omega \tau \rightarrow 0 $ ), the
first term does not contribute. It gives $ \omega^2 = \frac{1}{2} [ 3 
(\omega_{x}^{2} + \omega_{y}^{2}) \pm \sqrt{ 9 (\omega_{x}^{2} + \omega_{y}^{2} 
)^2 - 32 \omega_{x}^{2} \omega_{y}^{2}}] $.
This eigen frequencies exactly matches with Eq. (\ref{mono}), a result we 
 found using the equation of motion for the velocity fluctuations even
in deformed trap also. We have considered a few low energy 
excitation modes for which $ \nabla.\delta \vec v $ is constant.The first term of
 the right-hand side of Eq. (\ref{main}) does not contribute in the excitation spectrum.
Thats why the frequencies of these normal modes are same for a Bose gas just above
$ T_c $ and at very high temperature. 
 In pure collisionless regime ( $ \omega \tau \rightarrow \infty $ ), it 
gives $ \omega_{C} = 2 \omega_{x} $ and $ \omega_{C} = 2 \omega_{y} $.

 We can write phenomenological interpolation
formula for the frequency and the damping rate of the modes in the
following form  \cite{odelin}-\cite{pethick},
\begin{equation}
\omega^2 = \omega_{C}^2 + \frac{\omega_{H}^2 -  \omega_{C}^2}{1 -
i \omega \tau }
\end{equation}
The imaginary part of the above equation gives for the damping rate
\begin{equation}
\Gamma = \frac{\tau}{2}  \frac{d}{1 + ( \omega \tau )^2 }
\end{equation}
where $ d = (\omega_{C}^2 - \omega_{H}^2) $.
In the hydrodynamic limit ($ \omega \tau \rightarrow 0 $), the damping rate is 
\begin{equation}\label{damp1}
\Gamma_{HD} = \frac{\tau}{2} d
\end{equation}
where as in the collisionless region ($ \omega \tau \rightarrow \infty $),
\begin{equation}\label{damp2}
\Gamma_{CL} =  \frac{d}{2 \omega_{C}^2 \tau } 
\end{equation}
The damping rate depends on the difference between the square of the frequencies in the
collisional and hydrodynamical regime. The damping rates can be calculated for 
different values of temperature,  number of trapped atoms as well as of the 
trapping parameters and scattering length through the relaxation time $ \tau $ 
(\ref{tau}) .  For monopole mode in an isotropic trap the difference $ d $ 
 is zero.  So there is no damping in the monopole mode
in a 2D isotropic trapped Bose system when the tempareture is very high. It was first 
shown by  Boltzmann \cite{ford} and later Odelin et.al \cite{odelin} in a 3D trapped Bose
system at very high temperature.

For isotropic harmonic trap, Eqs. (\ref{bequ}) - (\ref{eequ}) decouples into two
subsystem, one is for monopole and other one for quadrupole mode.
The closed set of equations for monopole mode are
\begin{equation}\label{isobequ}
<\ddot{ \chi_{1}} > = 2< \chi_{5} > - 2 \omega_{0}^2 < \chi_{1} >
\end{equation}
\begin{equation}
<\ddot{ \chi_{3}} > = - 4 \omega_{0}^2 < \chi_{3} >
\end{equation}
\begin{equation}\label{isoeequ}
<\ddot{ \chi_{5}} > = 2 \omega_{0}^4 < \chi_{1}> + 2 \omega_{0}^2 
<\chi_{3}> 
\end{equation}
There is no collisional term in the above equations. So there is
no damping for the monopole mode of a classical dilute gas confined in a
isotropic trap. We are looking for solutions of Eqs. 
(\ref{isobequ}) - (\ref{isoeequ}) as 
$ e^{-i \omega t } $, we get $ \omega = 2 \omega_{0} $.

The Eqs. (\ref{isobequ}) - (\ref{isoeequ}) can be re-written as 
\begin{equation}\label{width}
\ddot{<\chi_{1}>} - \frac{1}{2 <\chi_{1}>} (\dot{<\chi_{1}>}^2) + 2
\omega_{0}^2 \chi_{1} = \frac{Q}{\chi_{1}}
\end{equation}
where $ Q = 2 (<\chi_{1}> <\chi_{5}> - <\chi_{3}>^2 ) $ is invariant
quantity
under time evolution.
Define $ X( t ) = \sqrt{<\chi_{1}>} $ which is the wave packet width and
substituting it into Eq. (\ref{width}) gives, 
\begin{equation}\label{hill}
\ddot X  + \omega_{0}^2 X = \frac{Q}{X^3}
\end{equation}
This is a nonlinear singular Hill equation. The same equation is
obtained
at $ T = 0 $ in 2D by Garcia Ripoll et.al \cite{garcia}.
At equlibrium, $ X_{0}^4 = \frac{Q}{ \omega_{0}^2 } $.
We linearized the Eq. (\ref{hill}) around the equilibrium point $ X_{0} 
$, we get
\begin{equation}
\ddot{\delta{X}} + 4 \omega_{0}^2 \delta X = 0
\end{equation}
One obtains an oscillation frequancy of the gas is
$ \omega = 2 \omega_{0} $, corresponding to the frequency of a single
particle excitation in the gas.

For time dependent trap, the equation of motion for the width of the wave packet is
\begin{equation}
\ddot X  + \omega_{0}^2(t) X = \frac{Q}{X^3}\label{hill1}
\end{equation}
The general solution  \cite{pin} is 
$ X(t) = \sqrt{u^2(t) + \frac{Q}{W^2} v^2(t)} $
where $ u(t)$ and $ v(t) $ are two linearly independent solutions of the equation 
$ \ddot p + \omega_{0}^2(t) p = 0 $ which satisfy $ u(t_0) = X(t_0) $,
$ \dot u(t_0) = X^{\prime}(t_0) $, $ v(t_0) = 0 $, $ v^{\prime}(t_0) \neq 0 $.
W is the Wronskian.This time dependent Hill equation (\ref{hill1})  can be solved 
explicitly only for a paricular choices of $ \omega_{0}(t) $.

The  closed set of equation for quadrupole mode in isotropic trap are
\begin{equation}
\ddot {<\chi_{2}>} = 2 <\chi_{6}> - 2 \omega_{0}^2 <\chi_2 >
\end{equation}
\begin{equation}
\ddot {<\chi_{4}>} = -4 \omega_{0}^2 <\chi_{4}> - \frac{<\chi_{6}>}{\tau }
\end{equation}
\begin{equation}
\ddot {<\chi_{6}>} =  2 \omega_{0}^4 <\chi_{2}> - \frac{<\dot \chi_{6}>} 
{\tau} - 2 \omega_{0}^2 <\chi_{6}>
\end{equation}
Solving this set of equation, we get damped quadrupole mode,
\begin{equation}
( \omega^2 - 4 \omega_{0}^2 ) + \frac{i}{\omega \tau} ( \omega^2 - 2
\omega_{0}^2 ) = 0
\end{equation}
In the hydrodynamic regime, the oscillation frequency is $ \omega^2 = 2
\omega_{0}^2 $ where as in the collisionless region, the frequency is just
a single particle oscillator frequency.
The damping rate can be calculated by using the Eqs. (\ref{damp1}) and (\ref{damp2}).

\section{SUMMARY }

In this work, we derived the equations of motion for velocity fluctuations of a 
Bose gas
in a 2D deformed trap potential just above the critical temperature. When $ U_{0} 
= 0 $, it 
becomes a wave equation, from which we found the exact sound velocity at high 
temperature. 
We have also computed the frequency of the scissors mode in hydrodynamic 
regime above $ T_{c} $ which agrees with the result obtained by Odelin et.al 
\cite{ode}. We have also calculated
the frequencies for monopole and quadrupole mode and the corresponding density 
fluctuations in a deformed trap above $ T_{c} $.

Using the method of averages, we obtained a  dispersion relation that 
interpolates between the collisionless and hydrodynamic regimes at very high 
temperature. 
In a deformed trap as well as an isotropic trap, we have found the frequencies 
and the damping rates (in terms of the relaxation time) for monopole and 
quadrupole mode  in both the regimes.
In hydrodynamical regime, the excitation frequencies for monopole and 
quadrupole mode exactly matches with the previous 
result that we have found from equatin (\ref{main}). We have also shown that the 
relaxation time $ \tau $ varies as $ \sqrt T $ in quasi-2D Bose gas.

We have shown that there is no damping for monopole mode in a 2D isotropic trapped
Bose gas when the temperature is very high. It was shown by Boltzmann \cite{ford}, later
Odelin et.al \cite{odelin} for 3D isotropic trap. 

We have discussed about the time evolution of the wave packet width of a Bose gas in a 
time independent as well as time dependent isotropic trap.
It can be described by the solution of the Hill equation.

\section{ACKNOWLEDGEMENT}
I would like to thank G. Baskaran and Subhasis Sinha for helpful discussions.

\end{document}